\documentclass[%
 reprint,
 amsmath,amssymb,
 aps,
]{revtex4-2}

\usepackage{aas_macros}

\usepackage{graphicx}% Include figure files
\usepackage{dcolumn}% Align table columns on decimal point
\usepackage{bm}% bold math
\usepackage{xcolor}

\begin{document}

\preprint{APS/123-QED}

\title{Correlation between planet formation rate and gas surface density: \\An analog of Kennicutt–Schmidt law
 for planet formation}%

\author{Mor Rozner}
 \email{morozner@ast.cam.ac.uk}
\affiliation{%
 Institute of Astronomy, University of Cambridge, Madingley Road, Cambridge CB3 0HA, UK\\
}%
\affiliation{
 Gonville \& Caius College, Trinity Street, Cambridge, CB2 1TA, UK\\
}%
\affiliation{
 Institute for Advanced Study, Einstein Drive, Princeton, NJ 08540, USA
}%

\date{\today}% It is always \today, today,

\begin{abstract} 
The efficiency of planet formation is a fundamental question in planetary science, gaining increasing significance as observational data from planet-forming disks accumulates. Here we derive from first principles a correlation between the planet formation rate (PFR) and the gas surface density, i.e. $\rm{PFR}\propto \Sigma_g^n$. This relation serves as an analog for the well-established Kennicutt-Schmidt law for star-forming galaxies. We study the different planet formation mechanisms and the density dependence in each one of them, to finally formulate a simple relation. We find that the powerlaw ranges between $n\approx 4/3-2$, depending on the type of the forming planet, when we carry out different analyses for the formation rates of terrestrial planets, gas giants, and also planets formed by gravitational instability. 
 We then compare our results with the available observational data. 
 The relation we derive here aims to shed more light on the interpretation of observational data as well as analytical models, and give a new perspective on the properties of planet formation and its connection to gas.
\end{abstract}

\maketitle

%\tableofcontents

\section{Introduction}

Planet formation is one of the most fundamental and enduring questions in astrophysics, spanning over a wide range of scales and processes -- from dust grains to full-scale planets, unfolding complex interplay of gravitational, hydrodynamic, and thermodynamic phenomena \citep{Armitage2020}. From the pioneering theories 
in the eighteenth century to the high-resolution observations of modern telescopes like ALMA \citep{ALMA2015}, our understanding of planet formation has evolved dramatically. Yet, despite significant advances, many aspects of this process are still under active research and the picture is far from being complete. 
 
Planet formation requires significant amounts of gas. The necessity of gas for the process of planet formation is manifested by the model of minimum mass solar nebula (MMSN) \citep{Weidenschilling1977,Hayashi1981}, which sets the minimal amount of material in the solar nebula to form the current Solar system.
In the time when planets are born, the protoplanetary disk is strongly dominated by gas ($99\%$), hinting at the significance of gas in the processes involved. 
The role played by gas during planet formation changes from one stage to another, but is essential for all of them, until the disk dispersal. At early stages, the gas determines the velocity of objects moving in it, setting the radial drift \citep{Weidenschilling1977}, and in late stages, gas drag-induced accretion determines the growth rate planetesimals \citep{Safronov1972,
OrmelKlahe2010,PeretsMurrayClay2011,LambrechtsJohansen2012}, and finally the growth of cores to gas giants \citep{PerriCameron1974,Mizuno1980}.  

Similar to planet formation, star formation also requires high gas densities.
Kennicutt-Schmidt law \citep{Kennicutt1989,Schmidt1959} is an empirical law that describes the relation between the star formation rate (SFR) and gas (surface) density, that now is shown to hold over a large number of star-forming galaxies \citep{Bigiel2008,Leroy2013,Pessa2021,Sun2023}.
Kennicutt-Schmidt law has a local (microscopic) version, specifies the regional SFR as a function of the gas surface density, 
as well as a global (macroscopic) one related to disk-averaged properties.
This law plays a crucial role in our understanding of the processes and conditions involved in star formation and also serves as a critical component in cosmological simulations. 

In this letter, we derive from first principles a relation connecting the planetary formation rate (PFR) to the gas surface density $\Sigma_g$ (or equivalently, the volumetric gas density $\rho_g$). We discuss a local PFR law, but relate also to a global version. Such a relation could be useful both for theoretical studies and future observations. There are some key differences between planet formation and star formation.
While stars are formed directly by the collapse of gas clouds, the dependence of planet formation on the surrounding gas is more indirect and complicated. Still, the existence and density of gas in a protoplanetary disk is a necessary component to enable planet formation.

\section{The final stages of planet formation}

Planet formation could be roughly divided into three stages: dust growth by coagulation, the intermediate regime and gravitationally-assisted growth. While the first and last stages are fairly understood, the intermediate one is still highly uncertain, as meter-sized objects should overcome various barriers to grow to km-sized objects, including the meter-size barrier, fragmentation, bouncing and aeolian-erosion \citep{Whipple1972,
Adachi1976,
Weidenschilling1977,BlumWurm2008,Zsom2010,
AeolianErosion}.

The final planet formation rate is determined essentially by the last stages before the gas disk dispersal, assuming a bottom-up growth of planets (see further discussion on gravitational instability in \ref{subsec:gi}). Hence, we will briefly review the growth rate of objects during these stages, and the relation to the gas density.
It is important to note that we are agnostic here to the planetesimal formation mechanism, and focus only on the growth after the planetesimal formation.  

\subsection{Terrestrial planets}

The growth of objects larger than planetesimals is governed by collisions of planetesimals/protoplanets either between themselves or with smaller objects, coupled to the gas -- pebble accretion, which is an efficient mechanism for protoplanets growth \citep{OrmelKlahe2010,PeretsMurrayClay2011,LambrechtsJohansen2012}. 
Pebble accretion describes the accretion of aerodynamically small particles on gravitationally large objects, and involves gas drag and gravity, with a typical rate of $\dot M_{\rm{PA}}\approx 2v_\infty b_{\rm{PA}} \Sigma_{\rm{peb}}$ where $v_\infty$ is the unperturbed velocity of the particle, $b_{\rm PA}$ is the impact factor and $\Sigma_{\rm peb}$ is the surface density of the pebbles. There are two regimes of pebble accretion based on the impact parameter, the shear regime -- large $b_{\rm {PA}}$ and headwind regime -- small $b_{\rm{PA}}$ \citep{Ormel2017}

\begin{align}
\dot M_{\rm PA} \approx \begin{cases}
\sqrt{8GM_{\rm {pl}}t_{\rm {stop}}v_{\rm {hw}}}\Sigma_{\rm {peb}}, \ \rm{headwind}, \\
2R_{\rm {Hill}}^2 \Omega_K \tau_s^{2/3} \Sigma_{\rm {peb}}, \ \rm{shear}
\end{cases}
\end{align}

\noindent
where $M_{\rm pl}$ is the mass of the accreting object, $t_{\rm stop}=mv_{\rm rel}/F_D$ is the stopping time of a particle with mass $m$, radius $R_{\rm peb}$ and a velocity $v_{\rm rel}$ relative to the gas, which applies a drag force of $F_D = 0.5 C_D \pi R_{\rm peb}^2 \rho_g v_{\rm rel}^2$, $C_D$ is a function of the Reynolds number \citep{PeretsMurrayClay2011}, $v_{\rm hw}$ is disk headwind, $\Omega_K$ is the Keplerian frequency of the particle and $\tau_s = \Omega_K t_{\rm{stop}}$.

The pebble surface density $\Sigma_{\rm peb}$ is determined by the pebble flux, and is changing in time, as it depends on the drifting particles. Overall \citep{LambrechtsJohansen2014}, 

\begin{align}
\Sigma_{\rm{peb}}=
\frac{\dot M_{\rm{peb,disk}}}{2\pi r v_{\rm{drift}}(\tau_s)}
=
f\Sigma_g \frac{r_{pro,0}}{r_{\rm{pro}}}\frac{v_{\rm{drift,0}}}{v_{\rm{drift}}}:=f_{\rm peb}\Sigma_g
\end{align}

\noindent
where $f=\Sigma_d/\Sigma_g$ is the dust-to-gas ratio, $r_{\rm{pro}}$ is the pebble production line, defined by the location, which drifts with time, at the disk in which the growth and drift timescale of a pebble are comparable, $v_{\rm{drift}}$ is the radial drift velocity \citep{Weidenschilling1977}, and $f_{\rm{peb}}=\Sigma_{\rm{peb}}/\Sigma_g$ is the pebbles-to-gas fraction. Subindex zero relates to the initial reference values, that change with time. We will adopt a typical value of $f_{\rm{peb}}=0.01$.

\subsection{Gas giants}

For some objects, the accretion is rapid enough, and they attain large masses before the gas disk dispersal. These objects will capture a significant gas envelope and will finally develop to become gas giants \citep{PerriCameron1974,Mizuno1980}. In this case, their formation rate will depend on the rate of the runaway gas accretion.
Here we adopt Bondi accretion rate, 

\begin{align}
\dot M_{\rm{RA}}\approx 4\pi R_{\rm Bondi}^2 c_s \rho_g 
\end{align}

\noindent
where $R_{\rm Bondi}=GM/c_s^2$ and $c_s$ is the sound speed in the disk. 
The gas accretion rate is then stopped/modified either when the gas supply from the disk is exhausted or the planet opens a gap \citep{TanigawaTanaka2016,GinzburgChiang2019}. Note that there could be corrections to the accretion rate depending on the thermal mass parameters, but they will not change steeply the overall scaling with the gas density \cite{ChoksiChiang2023}. For gap-opening planets, the mass accretion rate will become \citep{TanigawaWatanabe2002,TanigawaTanaka2016}

\begin{align}
\dot M_{GO}\approx 0.29 \left(\frac{h_p}{r_p}\right)^{-2}\left(\frac{M}{M_\star}\right)^{4/3}\Sigma_g r_p^2 \Omega_K
\end{align}

\noindent
where $h_p$ and $r_p$ are the scale height of the protoplanetary disk and the location of the planet correspondingly. $M_\star$ is the mass of the host star. 

\subsection{Other planet formation models}\label{subsec:gi}

Planet formation is not necessarily a bottom-up process. 
An alternative mechanism, known as gravitational instability (GI)
\citep{Kuiper1951,
Safronov1972,GoldreichWard1973}, 
 can lead to the formation of massive planets through instabilities in the protoplanetary disk.
GI bears similarities to the star formation process, enabling us to establish an analogous relation between the planetary formation rate (PFR) and the gas surface density,

\begin{align}
\rm{PFR}_{\rm{GI}}\propto \frac{\Sigma_g}{t_{\rm{ff}}}\propto \Sigma_g^{3/2}
\end{align}

\noindent
where $t_{\rm{ff}}=\sqrt{2\pi/32G\rho_g}$ is the free-fall timescale which is the typical timescale for instability to develop in the disk. The proportionality factor is determined by the efficiency of converting disk material into planets. 

\section{Kennicutt-Schmidt law for planet formation}

Using the growth rates we introduced above, we construct the PFR based on our current knowledge of planet formation stages. We define the PFR as a quantity that measures the number of planets, of a given type, that are forming in a region of the disk with a certain gas density.   
We estimate the PFR by

\begin{align}\label{eq:PFR}
\rm{PFR}\approx 
 \epsilon_{\rm PFR} \Sigma_p\Gamma_{\rm grow}\approx
 \epsilon_{\rm{PFR}}\frac{\Sigma_{\rm p}}{t_{\rm grow}}
\end{align}

\noindent
where $\epsilon_{\rm PFR}$ is an efficiency proportionality constant, that varies from one formation channel to another, $\Sigma_p$ is the surface density of "growing seeds", i.e. protoplanets for the case of terrestrial planets and cores for gas giants formation, $\Gamma_{\rm{grow}}\approx t^{-1}_{\rm{grow}}\approx M^{-1}\frac{dM}{dt}$ is the typical growth rate and $dM/dt$ is the mass accretion rate which we described above for the different regions. $t_{\rm{grow}}$ serves as the analog of the free-fall time, which dictates the typical timescale for star formation. The quantities $M, dM/dt$ are evaluated at typical seed masses of the forming planets, and are used to estimate the typical growth timescale $t_{\rm{grow}}$ (see Table \ref{table:fiducial} for fiducial values). We set the efficiency of protoplanets/cores formation by $\epsilon_{\rm{pro}}=\Sigma_{\rm{pro}}/\Sigma_{\rm{peb}}$ and $\epsilon_{\rm{core}}=\Sigma_{\rm{core}}/\Sigma_p$ correspondingly, and follow \citep{LambrechtsJohansen2014} to quantify them. They relate mainly to core formation, but we used that for protoplanet formation efficiency as well, as a restrictive value.
Overall, we find the following dependencies

\begin{align}
\rm{PFR}\propto \Sigma_g^n , \ n=\begin{cases}
3/2, \ \rm{terrestiral,headwind}\\
4/3, \ \rm{terrestiral,shear}\\
2, \ \rm{gas \ giants}\\
2, \ \rm{gap \ opening \ gas \ giants}\\
3/2, \ \rm{gravitational \ instability}
\end{cases}
\end{align}

\noindent
Similarly, we could derive a volumetric PFR law. It is interesting to note that while the formation processes of planets and stars are intrinsically different, we predict a dependence similar to the one initially derived for the SFR law \citep{Schmidt1959}, for two of the planet formation regimes. 

Observations of gas densities in protoplanetary disks as well as planet formation
face many non-trivial challenges. Most of the gas there is made of molecular hydrogen, which does not emit efficiently at low temperatures, such that observational data have to rely on different measurements \citep{Thi2001,Yoshida2022}. 
Currently, we found two protoplanetary disks that could be used to test our theory, in which the gas density and the planetary masses are known. Since the accretion rates on the planets are known in these disks, we will estimate the PFR by $\rm{PFR}_{\rm{obs}}\sim \dot M/A$
where $A$ is the surface
of the disk.
TW Hya is one of the most studied protoplanetary disks, with two major dust gaps, which are thought to host two super-Earths with a mass of $\sim 4 \ M_\oplus$ each \citep{Mentiplay2019,Yoshida2024}, and the mass accretion was found to be $4\times 10^{-7} -10^{-5} \ M_J \rm{yr}^{-1}$ \citep{Yoshida2024}, we will estimate $A\sim \pi (100 \rm{AU})^2$. 
Overall, $\rm{PFR}_{\rm{TH}}\approx 
1.9\times 10^{-11}-6.4\times 10^{-10}  
\ M_{J} \rm{yr}^{-1}\rm{AU}^{-2}$. The corresponding gas surface density is $\Sigma_g = 10-10^2 \ \rm{g} \ \rm{cm}^{-2}$ \citep{Calahan2021,Yoshida2024}. PDS 70 hosts two protoplanets, with masses of $2-4 \ M_J$ and $1-3 \ M_J$ and accretion rates of $3\times 10^{-7}-8\times10^{-7} \ M_J \rm{yr}^{-1}$ and $10^{-7}-5\times10^{-7} \ M_J \rm{yr}^{-1}$ correspondingly \citep{Wang2020}, we will estimate $A\sim \pi(100 \rm{AU})^2$
The gas surface density is taken to be $10^{-3}-0.1 \ \rm{g}\rm{cm}^{-2}$ \citep{Portilla-Revelo2023}.
Overall the PFR could be estimated by $1.3\times 10^{-11}-4.1\times 10^{-11} \ M_J \rm{yr}^{-1}\rm{AU}^{-2}$.

\begin{figure}
    \centering
    \includegraphics[width=1\linewidth]{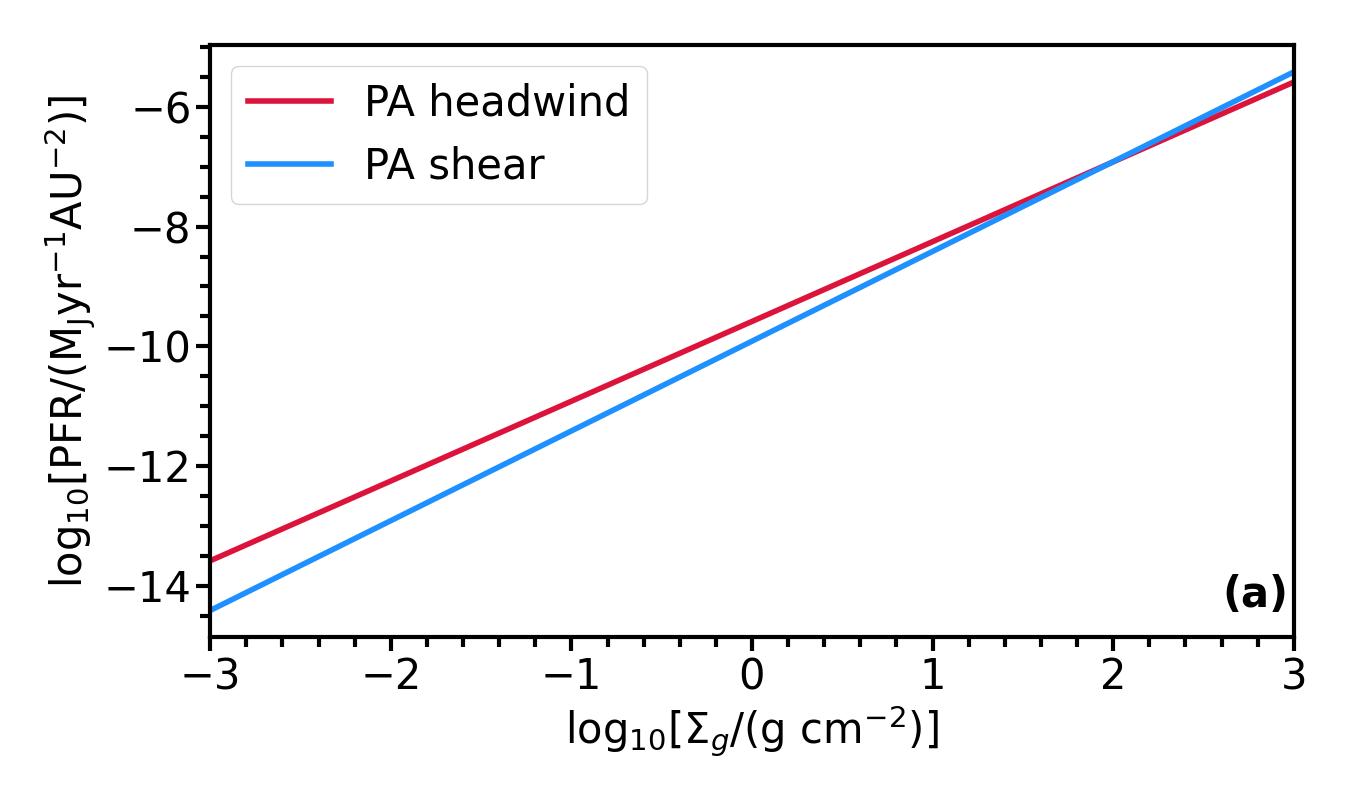}
    \includegraphics[width=1\linewidth]{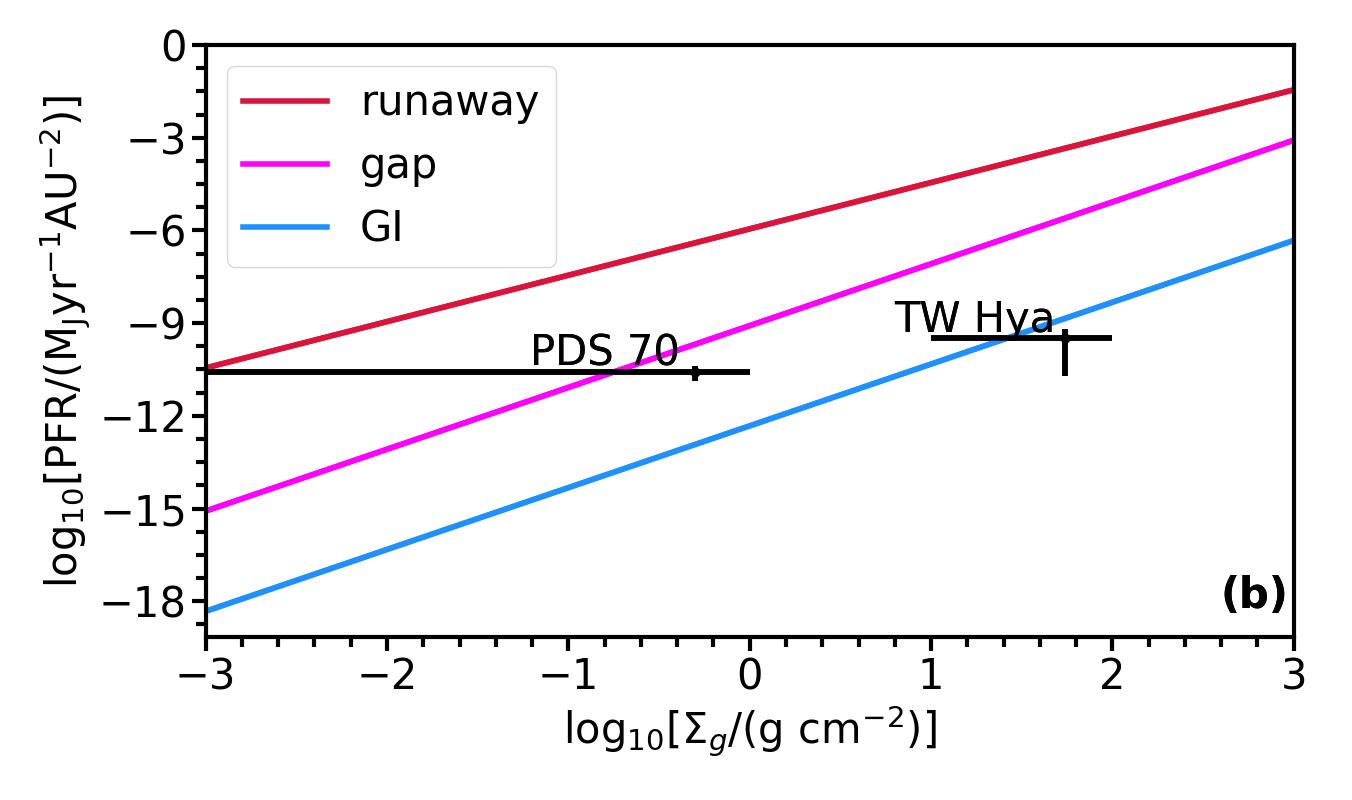}
    \caption{The planet formation rate (PFR) as a function of the gas surface density ($\Sigma_g$) for (a) terrestrial planets (b) gas giants. We consider different planet formation mechanisms (solid lines) and observational data (black crosses).}
\label{fig:PFR_Sigmag}
\end{figure}

In Fig. \ref{fig:PFR_Sigmag}, we present the dependence of the PFR on the gas surface density $\Sigma_g$ for different types of planets and formation processes. In Fig. \ref{fig:PFR_Sigmag}a, we consider the PFR for terrestrial planets, and in Fig. \ref{fig:PFR_Sigmag}b we consider the PFR for gas giants. As can be seen, in all the cases, there is a correlation between the PFR and the gas density. Moreover, the PFR is a monotonically increasing function of the gas density, indicating that there is positive feedback.
This trend could be seen also from the observational data we considered. 
It should be noted that the formation of terrestrial planets is more efficient than the formation of gas giants, as also concluded in \citep{Buchhave2012}.
In the near future, hopefully, we will have more observations on planet-forming disks that will enable us to test against larger statistics.  

\section{Discussion \& Summary}

In this letter, we derived from first principles a relation between the planet formation rate (PFR) and the gas surface density. Such a relation is an important step towards our understanding of planet formation, and hopefully will be tested against further observations in the near future, when statistics of planet-forming disks will be available. 

The law we derived here changes according to the planetary type, as different planets are formed via different physical processes. While we mostly focused on bottom-up formation, we also discussed the PFR for planets forming by gravitational instability. 
Our model could be used also to determine the dominant formation processes of planets, given the different dependencies. 

To extend the analogy to the SFR law introduced by \citep{Schmidt1959,Kennicutt1989}, one can define a threshold density for planet formation. Such a density was discussed in the context of the MMSN \citep{Weidenschilling1977,Hayashi1981}. The law derived in this letter will hold under the condition that $\Sigma_g\gtrsim \Sigma_{\rm {th}}$. 
Disks with smaller gas densities will lack the ability to produce planets, and accordingly will not obey the law we derived. 
For planets formed via GI, another threshold should be applied using the Safronov-Toomre stability criterion \citep{Safronov1970,Toomre1964}. We also expect to have self-regulation of planet formation, similar to star formation. As time goes by, the gas is depleted and accordingly, fewer planets are forming, while obeying the PFR-gas relation. 

Here we related mainly to a local PFR law, focusing on the local relation between PFR and gas surface density. Future studies can 
include spatially averaged planet formation rates over the disk, as well as a more accurate calculation of $t_{\rm{grow}}$, based on full integration. Moreover, this derivation relates to the initial properties of forming planets, which could later change with their evolution and could include migration and later accretion, disruptions, and mergers. 

Similar relations could also be derived for earlier stages of planet formation, i.e. pebble and planetesimal formation, using similar lines of thought. While pebble formation processes are well understood, the production of planetesimals is still an open question. Hence, the uncertainty of the derived powerlaws in this case will be higher, as well as the current observational data. Formation of in-situ moons should present similar dependencies on the gas surface density. 

 Given more observations, we would be able to determine which formation mechanism is more favorable, given the different dependencies for different mechanisms. We could also gauge the overall efficiency factors that encapsulate the efficiencies of planet formation rates.  
Another trace for the relation of planet formation rate to the gas density could be found in discussions on the dependence of planet formation and distribution on metallicities, as higher gas densities correlate with lower metallicities \citep{IdaLin2004,
Wyatt2007,Buchhave2012}. 

Planet formation in distorted disks relies on different properties of the disks, and could give rise to different formation rates. Since a decent fraction of protoplanetary disks are thought to be distorted, a more detailed analysis of the planet formation rate in these disks should be carried out in the future.

\bibliography{apssamp}% Produces the bibliography via BibTeX.

\appendix

\section{Fiducial values}

\widetext
\begin{tabular}
{ |p{1cm}|p{6cm}| p{2.5cm}| }
\hline
\multicolumn{3}{|c|}{Fiducial values} \\
\hline
  Symbol & Meaning &Fiducial value\\
\hline
\hline
$\rho_{\rm{peb}}$ & pebble internal density & $3 \ \rm{g} \ \rm{cm}^{-3}$  \\
$R_{\rm{peb}}$ & pebble radius & $1 \ \rm{cm}$\\
$C_D$ & drag constant (for pebbles) & $1$ \\
$v_{\rm{hw}}$ & typical headwind for pebbles & $50 \ \rm{m} \ \rm{sec}^{-1}$ \\
$h$ & disk scale-height (at $\sim 1 \ \rm{AU}$) & $0.1 \ \rm{AU}$\\
$M_\star$ & host star mass & $1 \ M_\odot$\\
$c_s$ & speed of sound & $200 \ \rm{m} \ \rm{\sec}^{-1}$\\
$f$ & $\Sigma_d/\Sigma_g$, dust-to-gas ratio & 0.01\\
$f_{\rm{peb}}$ & $\Sigma_{\rm{peb}}/\Sigma_g$, pebble-to-gas ratio & $0.01$\\
$\epsilon_{\rm PFR}$ & planet formation efficiency & $1$ \\
$\epsilon_{\rm{pro}}$ & protoplanet formation efficiency & $0.2$\\
$\epsilon_{\rm{core}}$ & core formation efficiency & $0.2$\\
$M_{\rm pro}$ & typical $M_{\rm {pl}}$ for a protoplanet & $10^{-3} \ M_\oplus$\\
$M_{\rm{core}}$ & typical $M_{\rm{pl}}$ for a core & $10 \ M_\oplus$\\
$M_{\rm{gap}}$ & typical $M_{\rm{pl}}$ for a gap-opening planet & $1 \ M_J$\\
\hline
\end{tabular}\label{table:fiducial}

\end{document}